\documentstyle[aps,multicol,epsfig,prl,floats]{revtex}

\begin{document}
\draft
\twocolumn[\hsize\textwidth\columnwidth\hsize\csname @twocolumnfalse\endcsname
\title{Two lifetimes and the pseudogap in the orbital magnetoresistance
of Zn--substituted La$_{1.85}$Sr$_{0.15}$CuO$_4$}
\author{A. Malinowski $^{1,2}$, A. Krickser $^1$, Marta Z. Cieplak
$^{1,2}$, S. Guha $^1$,\\
K. Karpi\'{n}ska $^2$, M. Berkowski $^2$, and P. Lindenfeld $^1$}
\address{$^1$ Department of Physics and Astronomy, Rutgers University,
Piscataway, NJ 08854, USA\\
$^2$ Institute of Physics, Polish Academy of Sciences, 02 668 Warsaw,
Poland}
\maketitle

\begin{abstract}
The effect of zinc doping on the anomalous temperature dependence of the
magnetoresistance and the Hall effect in the normal state was studied in
a series of La$_{1.85}$Sr$_{0.15}$Cu$_{1-y}$Zn$_y$O$_4$ films, with
values of
$y$ between zero and 0.12. The orbital magnetoresistance at high
temperatures
is found to be proportional to the square of the tangent of the Hall angle,
as predicted by the model of two relaxation rates, for all Zn--doped
specimens,
including nonsuperconducting films. The proportionality constant is equal to
13.7 $\pm$ 0.5 independent of doping. This is very different from the
behavior
observed in underdoped La$_{2-x}$Sr$_{x}$CuO$_4$ films where a decrease
of $x$
destroys the proportionality. In addition, the behavior of the orbital
magnetoresistance at low temperatures is found to be different depending on
whether $x$ is changed or $y$. We suggest that these differences reflect a
different evolution of the pseudogap in the two cases.

\end{abstract}


]


The anomalous normal--state transport properties of cuprate
superconductors
represent a major challenge on the way toward the understanding of the
physics of these materials. The assumption that there are two different
relaxation times \cite{and,col} is an attempt to explain anomalies
observed in the resistivity, Hall effect and magnetoresistance.
It predicts a simple relation between the orbital magnetoresistance and
the Hall angle ($\Delta\rho/\rho \propto  \tan^2\Theta_H$), which is
indeed observed, at least at high temperatures in optimally doped
cuprates \cite{harris}. One explanation relates
the behavior of the Hall coefficient to the opening of the pseudogap in
the normal state \cite{hwang}. Recent photoemission studies even point to
the possibility of two different pseudogaps, a low--energy pseudogap
related to the evolution of the Fermi surface into discontinuous Fermi
discs in the underdoped cuprates \cite{nor}, and a high--energy pseudogap,
possibly related to the magnetic interactions \cite{ino}.

Recently we described measurements of the orbital magnetoresistance
(OMR) and the Hall effect on a wide range of La$_{2-x}$Sr$_{x}$CuO$_4$
(LSCO) films, from $x$ = 0.048 with no superconductivity
down to 4 K, through the superconducting range, to $x$ = 0.275
with properties approach those of a normal metal \cite{fedor}. We found
that the predicted relation between the Hall angle and the
magnetoresistance is not followed except in optimally doped films
at temperatures above 100K. At lower temperatures there is a point of
inflection in the curve of the OMR as a function of $T$, below which the
OMR increases rapidly. The large positive OMR observed below the
inflection
point survives in the nonsuperconducting specimens, indicating that it
cannot be attributed solely to superconducting fluctuations as
originally
suggested \cite{harris,kimura}. The point of inflection is seen to move
to higher temperatures as $x$ decreases in the underdoped specimens, and
we have suggested that this feature may be related to the opening of a
pseudogap as the metal--insulator transition is approached.

To reach a fuller understanding we have made measurements of OMR and Hall
effect on films of La$_{1.85}$Sr$_{0.15}$Cu$_{1-y}$Zn$_y$O$_4$ with $y$
from zero to 0.12.
Superconductivity is absent when $y$ is greater than 0.055.
We find that the change of $y$ gives rise to a distinctly different
evolution of the OMR than a change of $x$. In particular, the inflection
point on the OMR curve does not shift with $y$, consistent with a
pseudogap--opening that is unaffected by a change of $y$.
Moreover, the proportionality between
the OMR and $\tan^2\Theta_H$ is followed for all specimens, including
nonsuperconducting films, with a proportionality constant which does not
change with $y$ and remains equal to the value reported previously for
zinc--free LSCO \cite{harris}. This unexpected result is easily
explained by the models which use two relaxation times
\cite{and,col,harris},
but is much more difficult to understand on the basis of more conventional
Fermi--liquid theories which
assume anisotropic relaxation rates \cite{car,stoj,hlub,ioffe}.

The $c$--axis oriented films, about 6000 {\AA} thick, were grown by
pulsed laser deposition on LaSrAlO$_4$ substrates. The values of $y$ are
those of the targets, but
have been shown to be the same as in the films \cite{ciep}.
The specimens for the present study were selected for their small
residual resistivity. Their dependence of the in--plane resistivity on
temperature is shown in Fig.~1(a). The inset shows the room--temperature
resistivity, $\rho_{RT}$, as a function of $y$. It exceeds by about 30\%
that of similar single crystals \cite{fuku}. However, while $y$ was
no greater than 0.04 in the single crystals, we are able
to reach values three times as high without any deterioration of the
film quality \cite{ciep}.

The films were patterned by photolithography and the wires soldered
with indium to evaporated silver pads.
Standard six--probe geometry was used to measure Hall
voltage and magnetoresistance simultaneously. The measurements were made
in magnetic fields up to 8 T, in longitudinal fields (parallel to the
$ab$--plane) and in transverse fields (perpendicular
to the $ab$--plane and to the current), with $T$ between 25 K and 300 K.
The temperature was stabilized to about 3 ppm, as described
previously \cite{fedor}.

\begin{figure}[ht]
\epsfig{file=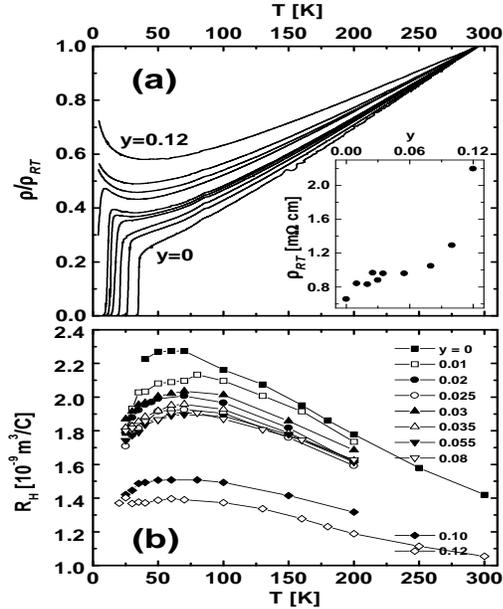, height=0.5\textwidth, width=0.5\textwidth}
\caption{(a) $T$--dependence of the resistivity, normalized to the
 room temperature value, $\rho_{RT}$, for films with various values
of $y$. From bottom up: 0, 0.01, 0.02, 0.025, 0.03, 0.035, 0.055,
0.08, 0.1, and 0.12. (b) $R_H$ as a function of $T$.}
\label{Fig.1}
\end{figure}

The Hall voltage is a linear function of field for all fields. Fig.~1(b)
shows the Hall coefficient, $R_H$, as a function of $T$. It is seen that
the increase of $y$ causes a decrease of $R_H$ without affecting the
shape of $R_H(T)$. The change in $R_H$ is about an order of magnitude
less than in LSCO when $x$ is decreased from optimal ($x$ = 0.15)
toward the strongly underdoped regime ($x=0.048$) \cite{hwang}.
The decrease of $R_H$ confirms the results previously observed in
ceramic specimens up to $y$ = 0.03 \cite{xiao}.
Note that a decrease of $x$ causes an increase of $R_H$ while an
increase in $y$ causes an opposite trend.
The change in $y$ does not lead to a superconductor--insulator
transition  \cite{fuku}, but rather to a metallic nonsuperconducting
phase \cite{mi,kasia}. The fact that the shape of $R_H(T)$
remains unaffected by the variation of $y$ is different from what
happens with overdoping of LSCO, which also leads to a metallic
phase, but destroys the anomalous $T$--dependence of $R_H$ \cite{hwang}.

Fig.~2 shows that the data for the Hall angle can be described by
$\cot\Theta_H = bT^2 + c$ from 25 K to 200 K. The variation of the
coefficients $b$ and $c$ is shown in the inset. The coefficient $c$
increases linearly with $y$ for all superconducting films at a rate
equal to 38 $\pm$ 4 per at.\% of Zn (and faster in nonsuperconducting
films). This is about three times as fast as
in zinc--substituted YBa$_2$Cu$_3$O$_{7-\delta}$ \cite{chien}.
The parameter $b$ is not constant, as suggested for
YBa$_2$Cu$_3$O$_{7-\delta}$ \cite{chien}, but increases with $y$.

\begin{figure}[ht]
\epsfig{file=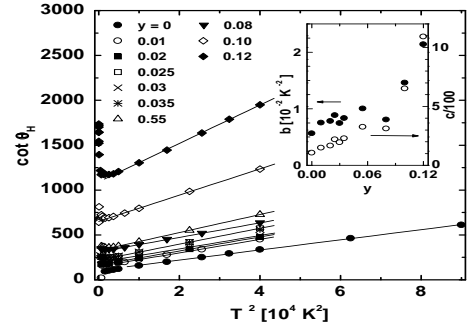, height=0.5\textwidth, width=0.5\textwidth}
\caption{The cotangent of the Hall angle at 8 T as a function of $T^2$.
The solid lines are fits to the relation $\cot \Theta_H = bT^2 + c$.
Inset: The coefficients $b$ and $c$ as functions of $y$.}
\label{Fig.2}
\end{figure}

The inset to Fig.~3 shows a typical example of the dependence of the
magnetoresistance on temperature. In all specimens the transverse
magnetoresistance (TMR) is positive down to 25 K, and it is always
larger than the longitudinal magnetoresistance (LMR). The LMR
is negative and very small above 200 K, approaching the
experimental resolution of the measurement. At lower T
the LMR becomes positive, and larger when $x$ decreases, or when
$y$ increases. In nonsuperconducting specimens with small $x$ or large
$y$ the LMR becomes negative and large below 25 K, consistent with the
expectation that the magnetic interactions, and the isotropic spin
scattering, which is presumably responsible for the LMR, then play an
increasingly important role \cite{artur}.

To obtain the OMR we subtract the longitudinal component from
the transverse magnetoresistance.
The temperature dependence of the OMR is shown in Fig.~3. A dramatic
suppression of the positive OMR occurs at low temperatures as $y$
increases, until it becomes negative in nonsuperconducting
specimens. This is very different from the behavior of the OMR in
LSCO, where a large positive OMR survives in nonsuperconducting films
\cite{fedor,artur}. This difference supports
our previous conclusion that superconducting fluctuations are not solely
responsible for the positive OMR observed at low temperatures. Since
superconducting fluctuations are expected to exist in the vicinity of
$T_c$ in underdoped and in zinc--doped LSCO, they would lead to the same
behavior in both types of specimens. Evidently this is not the case.

\begin{figure}[ht]
\epsfig{file=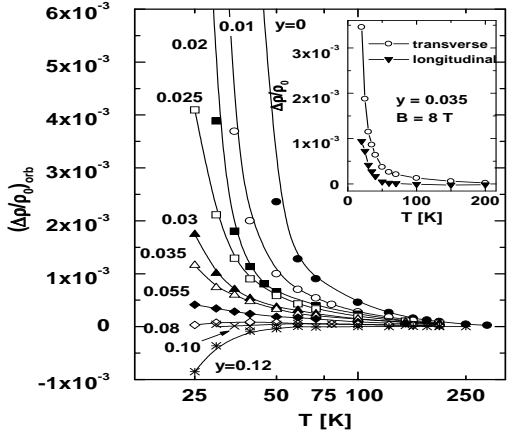, height=0.5\textwidth, width=0.5\textwidth}
\caption{The orbital magnetoresistance at 8 T as a function of $T$, for
all zinc-doped specimens. Inset: The temperature dependence of the
transverse and longitudinal magnetoresistance at 8 T for the film
with $y$ = 0.035. All lines are guides to the eye.}
\label{Fig.3}
\end{figure}
Fig.~4 shows the OMR on a logarithmic scale. It may be seen that the
point of inflection, which is at about 70 K in the film with $y=0$, does
not change its position along the $T$--axis with increasing $y$, while
with decreasing $x$ the point of inflection moves to higher temperatures.
We suggest that the fact that the shift is not observed in the
zinc--doped films indicates that the temperature
at which the pseudogap opens is not affected by the change of $y$.
This may be understood if one assumes that the zinc doping affects the
psedogap behavior only {\em locally}, in the immediate vicinity of the
impurity, but not
in the bulk of the specimen. Thus, while the doping affects the
scattering in the bulk of the specimen as seen by the fact that both the
Hall angle and the OMR above the point of inflection change with $y$,
the temperature of the pseudogap opening does not change. A similar
suggestion was made in a study of the thermopower in zinc--doped
YBa$_2$Cu$_3$O$_{7-\delta}$ and YBa$_2$Cu$_3$O$_8$ \cite{tallon}.

Further insight into the nature of the scattering comes from testing the
relation between the OMR and the square of the tangent of the Hall
angle. We find that this relation is followed for all
zinc--doped specimens at temperatures above the inflection point.
\begin{figure}[ht]
\epsfig{file=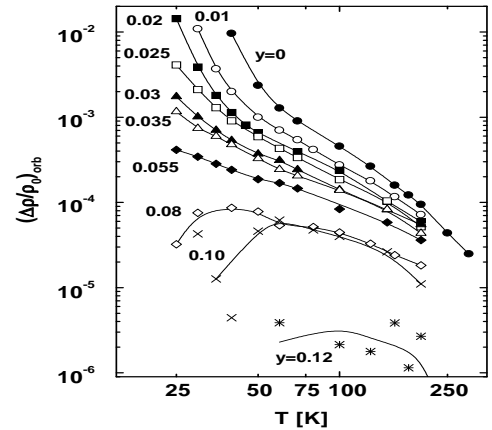, height=0.5\textwidth, width=0.5\textwidth}
\caption{The orbital magnetoresistance at 8 T as a function of temperature,
plotted on a log--log scale for all zinc--doped specimens.}
\label{Fig.4}
\end{figure}
Examples of this dependence are shown in the inset to Fig.~5 for three
films. The dotted lines are fits to the equation $a/(bT^2 + c)^2$.
A comparison of the experimental values of the OMR and $\tan^2\Theta_H$,
measured at temperatures above the inflection point for all of the
zinc--doped films, is shown on a log--log plot in Fig.~5. With the
exception of data for $y$ = 0.12, which are close to the limit of
resolution in our experiment, the data fall on straight
lines which have approximately the same slope. Small parallel shifts
between them probably result from experimental uncertainty of the sample
sizes. Excluding the data for $y$ = 0.12, we can fit the data with a
straight line with slope 0.94 $\pm$ 0.06. The proportionality constant $a$,
averaged over all data, is equal to 13.7 $\pm$ 0.5,
in excellent agreement with the value 13.6 reported for LSCO with
$x$ = 0.17 by Harris et al. \cite{harris}.

The observation that the proportionality constant $a$ is unaffected by
doping puts strong constraints on the theoretical models which attempt
to explain the anomalous properties of the normal state in cuprates.
These models may be divided into two classes. Those in one class, the
Fermi--liquid models, are based on the assumption that some strong,
unusual anisotropy of the relaxation rates around the Fermi surface
leads to the anomalies \cite{car,stoj,hlub,ioffe}.
Although details of these models vary, it would be expected that the
ratio of the OMR to $\tan^2\Theta_H$ would depend on temperature and
doping so that our observation would require some
fortuitous cancellation. The models in the second class assume the
existence of
two different relaxation rates at all points of the Fermi surface
\cite{and,col,harris}. It is a fundamental property of these models that
the ratio of the OMR to $\tan^2\Theta_H$ is constant, and should not be
affected by doping. These models thus appear to be favored by
our results. Finally we note that high--magnetic--field studies
of the magnetoresistance in Tl$_2$Ba$_2$CuO$_{6+\delta}$ also favor the
two lifetime models \cite{tyler}. Their microscopic understanding
remains, however,
elusive.
\begin{figure}[ht]
\epsfig{file=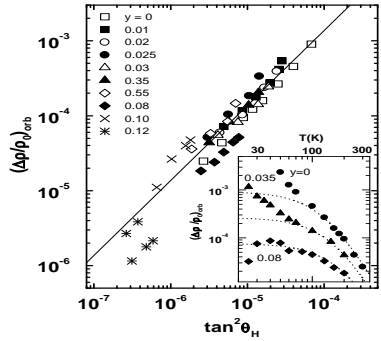, height=0.5\textwidth, width=0.5\textwidth}
\caption{
Log--log plot of the orbital magnetoresistance versus $\tan^2 \Theta_H$,
measured at temperatures above the inflection point in films with
various values of $y$. Inset: The orbital magnetoresistance at 8 T as
a function of temperature for three films with $y$ = 0, 0.035, and 0.08.
The dotted lines follow the equation $a/(b T^2+c)^2$.}
\label{Fig.5}
\end{figure}
We conclude that the metallic phase created by zinc doping retains
the anomalous characteristics that are observed at high temperatures in
optimally doped LSCO, in both the OMR and
the Hall effect, together with the relationship between them. We suggest
that the striking contrast between this result and our previous
observation of the dissapearance of the proportionality between OMR and
Hall angle in underdoped LSCO \cite{fedor} is related to the opening of
a pseudogap. Apparently the opening of the pseudogap destroys this
characteristic feature of the anomalous normal state.

We would like to thank M. Gershenson for his cooperation and sharing of
laboratory facilities, and Piers Coleman and Andrew Millis for
helpful discussions. We also thank Richard Newrock and the Physics
Department of the University of Cincinati for help with the construction
of the target chamber. This work was supported by the Polish Committee
for Scientific Research, KBN, under grant 2 P03B 09414, by the Naval
Research Laboratory, and by the Rutgers Research Council.

\end{document}